\documentclass[12pt]{article}
\usepackage{epsfig}

\begin{document}
%

\begin{titlepage}

\begin{flushright}
Fermilab-Pub-04-305-E\\
October 27$^{st}$ 2004            \\
\end{flushright}

\vspace*{2cm}

\begin{center}
\begin{Large}

{\bf  Measurement of the Top Quark Mass In All-Jet Events }

\vspace{2cm}

D\O\ Collaboration \\[0.5cm]

\end{Large}

\end{center}

\vspace{2cm}

\begin{abstract}

We describe a measurement of the mass of the top quark 
 from the purely hadronic decay modes of $t\overline{t}$ pairs 
using all-jet data
 produced in $p\overline{p}$ collisions
at $\sqrt s$ = 1.8 TeV 
 at the Fermilab Tevatron Collider.   
The data, which correspond to an integrated luminosity of
110.2 $\pm$ 5.8 pb$^{-1}$, were collected 
with the D\O~detector from 1992 to 1996.
We find a  top quark mass of 
178.5 $\pm$ 13.7 ~(stat)~$\pm$ 7.7~(syst)~GeV/$c^2$.

\end{abstract}


\vspace{1.5cm}

\begin{center}
Submitted to Physics Letters B
\end{center}

\end{titlepage}

\begin{flushleft}
%
\begin{center} 
V.M.~Abazov,$^{21}$                                                           
B.~Abbott,$^{54}$                                                             
A.~Abdesselam,$^{11}$                                                         
M.~Abolins,$^{47}$                                                            
V.~Abramov,$^{24}$                                                            
B.S.~Acharya,$^{17}$                                                          
D.L.~Adams,$^{52}$                                                            
M.~Adams,$^{34}$                                                              
G.D.~Alexeev,$^{21}$                                                          
A.~Alton,$^{46}$
% ...
\\                                                                            
\vskip 0.30cm                                                                 
\centerline{(D\O\ Collaboration)}                                             
\vskip 0.30cm 
\end{center}                                                                
\centerline{$^{1}$Universidad de Buenos Aires, Buenos Aires, Argentina}       
\centerline{$^{2}$LAFEX, Centro Brasileiro de Pesquisas F{\'\i}sicas,         
                  Rio de Janeiro, Brazil}                                     
\centerline{$^{3}$Universidade do Estado do Rio de Janeiro,                   
                  Rio de Janeiro, Brazil}                                     
\centerline{$^{4}$Institute of High Energy Physics, Beijing,                  
                  People's Republic of China}                                 
\centerline{$^{5}$Universidad de los Andes, Bogot\'{a}, Colombia}             
\centerline{$^{6}$Charles University, Center for Particle Physics,            
                  Prague, Czech Republic}                                     
\centerline{$^{7}$Institute of Physics, Academy of Sciences, Center           
                  for Particle Physics, Prague, Czech Republic}               
\centerline{$^{8}$Universidad San Francisco de Quito, Quito, Ecuador}         
\centerline{$^{9}$Laboratoire de Physique Subatomique et de Cosmologie,       
                  IN2P3-CNRS, Universite de Grenoble 1, Grenoble, France}     
\centerline{$^{10}$CPPM, IN2P3-CNRS, Universit\'e de la M\'editerran\'ee,     
                  Marseille, France}                                          
\centerline{$^{11}$Laboratoire de l'Acc\'el\'erateur Lin\'eaire,              
                  IN2P3-CNRS, Orsay, France}                                  
\centerline{$^{12}$LPNHE, Universit\'es Paris VI and VII, IN2P3-CNRS,         
                  Paris, France}                                              
\centerline{$^{13}$DAPNIA/Service de Physique des Particules, CEA, Saclay,    
                  France}                                                     
\centerline{$^{14}$Universit{\"a}t Freiburg, Physikalisches Institut,         
                  Freiburg, Germany}                                          
\centerline{$^{15}$Panjab University, Chandigarh, India}                      
\centerline{$^{16}$Delhi University, Delhi, India}                            
\centerline{$^{17}$Tata Institute of Fundamental Research, Mumbai, India}     
\centerline{$^{18}$CINVESTAV, Mexico City, Mexico}                            
\centerline{$^{19}$FOM-Institute NIKHEF and University of                     
                  Amsterdam/NIKHEF, Amsterdam, The Netherlands}               
\centerline{$^{20}$University of Nijmegen/NIKHEF, Nijmegen, The               
                  Netherlands}                                                
\centerline{$^{21}$Joint Institute for Nuclear Research, Dubna, Russia}       
\centerline{$^{22}$Institute for Theoretical and Experimental Physics,        
                   Moscow, Russia}                                            
\centerline{$^{23}$Moscow State University, Moscow, Russia}                   
\centerline{$^{24}$Institute for High Energy Physics, Protvino, Russia}       
\centerline{$^{25}$Lancaster University, Lancaster, United Kingdom}           
\centerline{$^{26}$Imperial College, London, United Kingdom}                  
\centerline{$^{27}$University of Arizona, Tucson, Arizona 85721}              
\centerline{$^{28}$Lawrence Berkeley National Laboratory and University of    
                  California, Berkeley, California 94720}                     
\centerline{$^{29}$California State University, Fresno, California 93740}     
\centerline{$^{30}$University of California, Irvine, California 92697}        
\centerline{$^{31}$University of California, Riverside, California 92521}     
\centerline{$^{32}$Florida State University, Tallahassee, Florida 32306}      
\centerline{$^{33}$Fermi National Accelerator Laboratory, Batavia,            
                   Illinois 60510}                                            
\centerline{$^{34}$University of Illinois at Chicago, Chicago,                
                   Illinois 60607}                                            
\centerline{$^{35}$Northern Illinois University, DeKalb, Illinois 60115}      
\centerline{$^{36}$Northwestern University, Evanston, Illinois 60208}         
\centerline{$^{37}$Indiana University, Bloomington, Indiana 47405}            
\centerline{$^{38}$University of Notre Dame, Notre Dame, Indiana 46556}       
\centerline{$^{39}$Iowa State University, Ames, Iowa 50011}                   
\centerline{$^{40}$University of Kansas, Lawrence, Kansas 66045}              
\centerline{$^{41}$Kansas State University, Manhattan, Kansas 66506}          
\centerline{$^{42}$Louisiana Tech University, Ruston, Louisiana 71272}        
\centerline{$^{43}$University of Maryland, College Park, Maryland 20742}      
\centerline{$^{44}$Boston University, Boston, Massachusetts 02215}            
\centerline{$^{45}$Northeastern University, Boston, Massachusetts 02115}      
\centerline{$^{46}$University of Michigan, Ann Arbor, Michigan 48109}         
\centerline{$^{47}$Michigan State University, East Lansing, Michigan 48824}   
\centerline{$^{48}$University of Nebraska, Lincoln, Nebraska 68588}           
\centerline{$^{49}$Columbia University, New York, New York 10027}             
\centerline{$^{50}$University of Rochester, Rochester, New York 14627}        
\centerline{$^{51}$State University of New York, Stony Brook,                 
                   New York 11794}                                            
\centerline{$^{52}$Brookhaven National Laboratory, Upton, New York 11973}     
\centerline{$^{53}$Langston University, Langston, Oklahoma 73050}             
\centerline{$^{54}$University of Oklahoma, Norman, Oklahoma 73019}            
\centerline{$^{55}$Brown University, Providence, Rhode Island 02912}          
\centerline{$^{56}$University of Texas, Arlington, Texas 76019}               
\centerline{$^{57}$Rice University, Houston, Texas 77005}                     
\centerline{$^{58}$University of Virginia, Charlottesville, Virginia 22901}   
\centerline{$^{59}$University of Washington, Seattle, Washington 98195} 
\vspace*{0.5cm}

\centerline{[*]     Visitor from University of Zurich, Zurich, Switzerland}

\centerline{[$\dag$] Visitor from Institute of Nuclear Physics, Krakow, Poland}      

\end{flushleft}

\newpage
\normalsize

\vfill\eject

The mass of the top quark ($m_t$) is a key parameter of the
standard model (SM). Knowledge of its  value is essential for determining
quantum corrections to the theory and for limiting the predicted
range of Higgs boson masses~\cite{quigg,ewstudies}. In this letter we report a new
measurement of $m_t$  by the D\O\ collaboration
in the process  $p \bar p \rightarrow t \bar t$
for the case in which the top and antitop quarks  each decay
to a $W$ boson and a $b$ quark, followed by the hadronic decay of both $W$ bosons.
At lowest order in perturbative QCD, this leads to a final state
of six quark jets, referred to as the all-jets channel of
$t \bar t$ production.

Measurements of $m_t$ have been reported by the D\O~and CDF
collaborations based on Run I data (1992--1996) from the Fermilab Tevatron
Collider, with a $p\bar{p}$ center-of-mass energy of 1.8 TeV.
 For the all-jets final state, CDF  has
reported  a top quark mass of 186 $\pm$ 10 (stat) $\pm$ 12 (syst)
GeV/$c^2$~\cite{cdfalljets}.
 CDF, combining leptonic and hadronic $W$ boson decay channels, has measured an  
 average top quark mass of
175.9~$\pm$ 4.8~(stat) $\pm$ 4.9~(syst) GeV/$c^2$~\cite{cdfmass}.
 The D\O~  average for $m_t$, based on
leptonic decay channels of one or both of the $W$ bosons,
is 179.0 $\pm$ 3.5 (stat) $\pm$ 3.8 (syst)  GeV/$c^2$~\cite{nature}.
Combining all published measurements yields an average value of
178.0 $\pm$ 4.3 GeV/$c^2$ for the mass of the top quark~\cite{nature}.

 The all-jets decay  channel  has the largest branching fraction
of all $t\bar{t}$  decay channels (46\%) and
is the most kinematically constrained
final state, since no energetic neutrinos are produced~\cite{PDG}. If 
the jets could be correctly associated
with their original partons, there would be no
ambiguity in the analysis. However, the association cannot be
made unambiguously.
The events of interest  contain
six or more high transverse momentum jets,
two of which originate from $b$ quarks.  
The dominant background arises from
other QCD processes that produce six or more jets.  
 D\O\ measured the $t\overline{t}$~production 
cross section in 
the all-jets decay channel to be 7.1   $\pm$ 2.8 (stat) $\pm$ 1.5 (syst)  pb,
assuming a  top quark mass of 172.1 GeV/$c^2$ that 
was previously determined by D\O~ from
 other decay channels~\cite{dzeroalljets}. 
This cross section corresponds to  roughly 360 $t\bar{t}\rightarrow jets$ 
events produced at D\O\ during Run I.

The D\O~detector and our  methods of triggering, identifying particles,
and reconstructing events are described elsewhere~\cite{dzerodet,dzero95}.  
The measurement reported here is based on 110.2~$\pm$ 5.8 pb$^{-1}$ of data
from Run I of the Fermilab Tevatron
Collider, at a $p\bar{p}$ center-of-mass energy of 1.8 TeV.
The events  were
selected with a trigger that required at least five jets 
of cone radius  $\mathcal R  \equiv  \sqrt{(\Delta \eta)^2 + (\Delta \phi)^2} = 0.3$,
where  $\phi$ and $\eta$ are the azimuthal angle
and pseudorapidity, respectively. Each jet was required to have 
transverse energy ($E_T$) greater than 10 GeV and $|\eta| < 2.5$. The 
triggered sample contains approximately 2 million events, with 
an estimated
signal-to-background ratio of about 1 to 7000. 
Further selection criteria  were applied to this sample,    
including criteria to suppress Main Ring noise, to ensure
good jet energy resolution, and to remove events 
consistent with light-quark backgrounds.
Events containing a reconstructed 
electron or muon  outside a jet cone of radius
$\mathcal R = 0.5$  were excluded to avoid  overlap with other  
$t\overline{t}$~ decay channels.  
In addition, we required
events to contain at least six  jets
with $E_T > 10$ GeV and $|\eta| < 2.5$.
The $E_T$ of each jet was scaled to give, on average, 
the correct (true) jet energy.
For more information on
the D\O\ jet algorithm, see Ref.~\cite{dzerojets}.
For information on the jet energy scale correction see Ref.~\cite{dzerocal}

These criteria led to  the selection of  a sample containing
165,373 multijet events with an estimated signal-to-background ratio of about
1 to 1000. 
Since  $t\overline{t}$ events  contain a $b\bar{b}$ pair, 
 whereas such pairs are relatively rare in background events, the
signal-to-background ratio can be improved by selecting events with at
least one jet that may have arisen from
the fragmentation of a $b$ quark that subsequently decayed to a muon,
either directly, or indirectly  via the chain $b \rightarrow  c \rightarrow \mu$.
The tagging of $b$-jets using embedded muons is
effective at suppressing background relative to signal because
15--20\% of $t\bar{t}$ events are so tagged 
whereas  only 2\% of the multijet events in the selected sample satisfy
the $b$-tagging requirements.
In this analysis, 
a $b$-jet was defined as any $\mathcal R$ = 0.5 cone jet, with
$E_T > 10$ GeV and $|\eta| < 1.0$, that contained a
muon with $p_T > 4$ GeV/$c$ within its cone.
The tag  requirement reduced the
sample to 3,043 $b$-tagged events, with an estimated 
signal-to-background ratio of approximately 1 to 100.
More details on $b$ tagging are given in Ref.~\cite{dzeromass}.

To determine the properties of  all-jets events from $t\overline{t}$~
production, extensive Monte Carlo  studies were done. The
$t\overline{t}$~ signal events were generated using the
{\sc herwig} ~V5.7 ~\cite{HERWIG}  
program and propagated 
through a detailed detector simulation, 
based on {\sc geant} ~V3.15 \cite{GEANT}, 
and reconstructed with 
the standard D\O\ reconstruction program. 
We found that
the mean of the invariant masses of two triplets of jets
formed from the six highest-$E_T$ jets,
$ M \equiv \left(\frac{m_{t_1}+m_{t_2}}{2}\right )  $,
provided a satisfactory discriminant for distinguishing 
$t \bar t$ signal from background ~\cite{thesis}. We chose
the two jet triplets to be
those that minimized the quantity
\begin{eqnarray}
\chi^2 & = & \left(\frac{m_{t_1}-m_{t_2}}{2\times \sigma_{m_t}}\right )^2 
        +  \left(\frac{M_{W_1}-M_{W_0}}{\sigma_{M_W}}\right )^2 
+ \left(\frac{M_{W_2}-M_{W_0}}{\sigma_{M_W}}\right )^2, 
\label{equation:chi}
\end{eqnarray}
where $M_{W_0} = 77.5$ GeV/$c^2$ is
the mean value of the reconstructed $W$ boson mass in the
 all-jets $t\overline{t}$  Monte Carlo events processed through 
the D\O\ detector simulation and reconstruction programs,
and $m_{t_1}$, $m_{t_2}$ and $M_{W_1}$, $M_{W_2}$ are
the calculated masses of the reconstructed
jets that correspond to candidate top quarks and $W$ bosons, respectively, 
computed from the jet triplets and, within each triplet, the jet doublets.
The standard deviations, $\sigma_{m_t} \approx 31$ GeV/$c^2$ 
and $\sigma_{M_W} \approx 21$ GeV/c$^2$,
are the average root mean square (RMS) values of the mass  
distributions determined using {\sc  herwig} Monte Carlo events
generated with top quark masses of 140, 180 and 220 GeV/$c^2$.  
Minimizing $\chi^2$ provides the correct combination of jets  in about 40\%
of the $t\overline{t}$~ Monte Carlo events.

The top quark mass was measured through the best fit  
of different admixtures of  
signal and background to the observed mass distribution.
The fitting technique used  
is similar to that of Ref.~\cite{BPS}, which
takes account of the finite size of every sample in the fit. The
posterior probability density $p(m_t, \sigma_{t\bar{t}}|\mbox{Data})$, 
computed 
assuming a flat prior in mass and in the $t\overline{t}$~ cross section,
is calculated for  
a set of mass values $m_t$.
For each $m_t$ value, the posterior  probability  density,
numerically identical (in this case) to the
likelihood $L$, was maximized by varying $\sigma_{t\bar{t}}$
to give the ``maximized likelihood'', $L_{max}(m_t)$
as a function of the hypothesized top quark mass, $m_t$. 
The ``best fitted mass'', $m_{fit}$, was taken to be the location
of the minimum of the negative log-likelihood curve
$-\ln{L_{max}(m_t)}$.

The templates for the top quark 
signal were generated using a  Monte Carlo simulation
of $t\bar{t}$ events for 
a discrete set of masses in the
range of 110 to 310~GeV/$c^2$ in 10 GeV/$c^2$ steps, that is, at 21 
mass values.  
The background
was modeled using {\em untagged} events, that is, multijet data that passed all
selection criteria except those that define the $b$-tag. 
For each untagged event $i$, a weight $w_i$
is calculated, which reflects the
probability of tagging that event, such that the
sum $\sum_i w_i$ over all untagged events provides an estimate of
the background, that is, the number
of non-$t\bar{t}$ $b$-tagged events within the 3,043 event sample.
The event weight is the sum of the
$b$-tag rate {\em per jet}, which is assumed to depend 
only on the  $E_T$ and $\eta$ of the jet, and on the muon detection 
efficiency. The tag-rate ($t_R$) is assumed to factorize as follows:
\begin{equation}
 t_R = T(E_T,\eta,R) = N(R)f(E_T,R)g(\eta,R) \, ,
\label{equation:tagrate}
\end{equation}
where 
\begin{equation}
f(E_T,R) = a_0+a_1E_T^{1/2} \, ,
\end{equation}
\begin{equation}
g(\eta,R) = p_0+p_1 | \eta | ^2             
\end{equation}
and N(R) is an overall normalization constant.
The forms of $f(E_T,R)$ and $g(\eta,R)$ were determined empirically.    
The tag-rate
is divided into 5 bins as a function of run number (R) according to
the 5 major changes in  muon system efficiency. 
For each of the 5 run bins, 
jets were selected with $E_T>$10 GeV and $| \eta | \leq $1.0.
Histograms were made of the $E_T$ of the tagged jets   and the $E_T$ of the 
untagged jets in the data, and their
ratio (bin by bin) was  fit using $f(E_T,R)$.
Similarly, the distributions of the ratio of tagged and 
untagged $|\eta |$  histograms were fit using
 $g( \eta ,R)$.
The tag-rate was normalized to return the number of 
observed tagged events in the data. $\chi ^{2}$ tests show very 
good agreement between the tagged data and 
the predicted background.
 More details on this method are given in Ref.~\cite{thesis}.

\begin{figure}[hbt]
\vspace*{-0.5cm}
\centerline{\epsfxsize=9.0cm\epsfysize=9.0cm\epsfbox{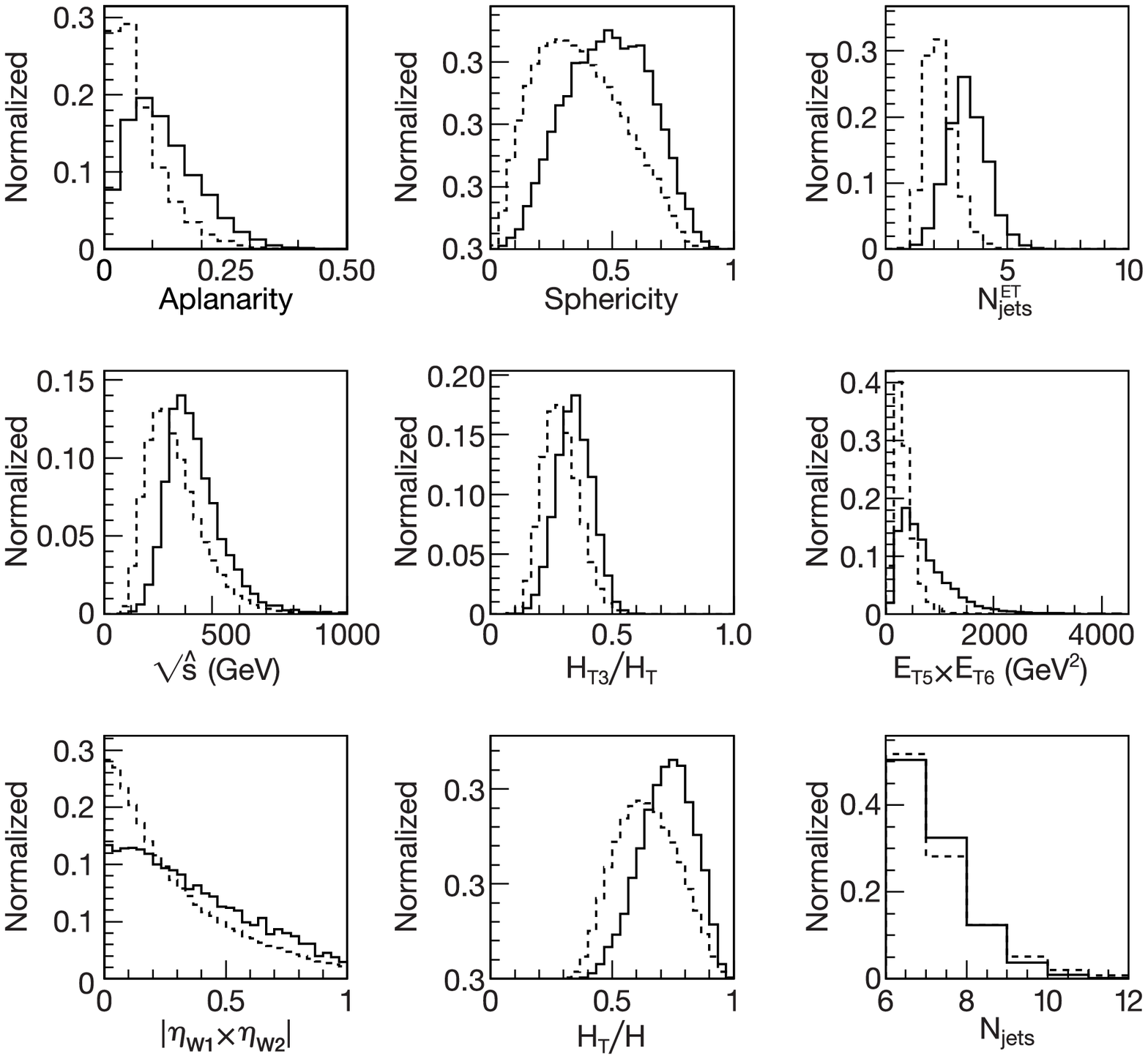}}
\caption{Histograms of  kinematic
variables for $t\overline{t}$~ Monte  Carlo  signal (solid) and
background (dashed) normalized to the same area. 
The Monte Carlo signal
samples were generated with a top quark mass of 180~GeV/$c^2$.
The variables are described in the text.}
\label{fig:neuralvars}
\end{figure}

Since the jets in $t\bar{t}$ events 
  tend to be more energetic, have a more isotropic
momentum flow, and have larger transverse energies than those
in light-quark events, we can enrich the event sample further 
by event discrimination based on a  suitable set of kinematic variables. 
For this analysis, we used the following
eight variables: $E_{T5} \times E_{T6}$, 
 $|\eta_{W1} \times \eta_{W2}|$, $\sqrt{\hat s}$,
${\cal A}$, ${\cal S}$, $N^{E_T}_{jet}$,
 $H_{T3}/H_T$, and $H_T/H$,  where $E_{T1}$ to $E_{T6}$ and $E_1$ to
$E_6$ are the
transverse energies and energies, respectively, 
of the six jets, ordered in decreasing $E_{T}$; $\eta_{W1}$ and $\eta_{W2}$ are the
pseudorapidities of the two hypothesized $W$ bosons; 
$\sqrt{\hat s}$ is the invariant mass of the $N_{jets}$ system;
${\cal A}$, the aplanarity, is  $\frac{3}{2}$ of
the smallest eigenvalue of the normalized laboratory-frame momentum
tensor~\cite{collider_physics} of all the jets; 
${\cal S}$, the sphericity, is  $\frac{3}{2}$ of the sum of the smallest
and next-smallest eigenvalues of the same tensor; $N^{E_T}_{jet}$ is the 
number of jets above a given $E_T$ threshold, over the range 10 GeV to 55 GeV, 
weighted by the threshold \cite{Tkachov}; 
$H_T = \sum_j E_{Tj}$; $H_{T3} = H_T - E_{T1} - E_{T2}$; and
$H = \sum_j E_j$, 
where the sums are over all ${\cal R} = 0.5$ cone jets with 
$|\eta| < 2.5$ and $E_T > 10$ GeV.
Figure~\ref{fig:neuralvars} shows comparisons of distributions in each
of the kinematic variables between  background and a 
$t\overline{t}$~ Monte Carlo signal for $m_t =$ 180~GeV/$c^2$. 
  The distributions of kinematic 
 variables for events with $b$-jets are consistent with those 
without $b$-jets.

The above variables were 
combined into a single discriminant, 
calculated using a neural network (NN)~\cite{NN}
with eight inputs, a single hidden layer with three nodes, 
and a single output $D_{NN}$. The network
was  trained and tested with independent samples
of {\sc herwig} Monte Carlo $t\overline{t}$~ signal events, 
at top quark masses of 140, 180 and 220 
GeV/$c^2$,
and untagged events for the background, with the target for
background set at 0 and at 1 for the signal. Roughly equal numbers of training
events were used at each mass. 
These events (in all,
11,423 for signal and 8,143 for background)
were used only for training the neural network and producing a
single set of network parameters. These events were not used 
in the subsequent analysis.

\begin{figure}
\epsfxsize=9.0cm
\centerline{\epsffile{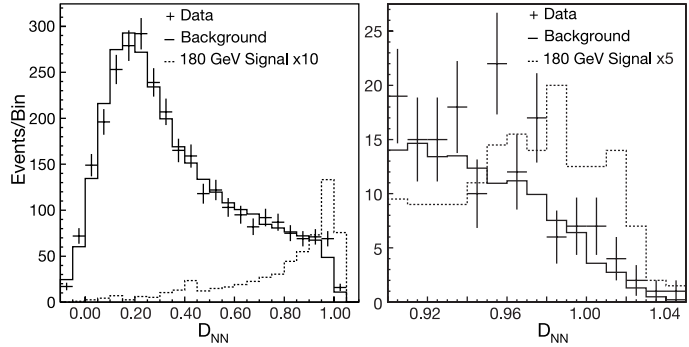}}
\caption{(a) $D_{NN}$ is plotted for data and background. Also shown  
is the $D_{NN}$ expected for a 180
GeV/$c^2$ top quark Monte Carlo
signal scaled up by a factor of ten. (b) $D_{NN}$ in finer bins from 0.90
to 1.05, for data, background and 180 GeV/$c^2$ top quark Monte Carlo
signal scaled up by a factor of five.
}
\label{fig:nn_data_bkg}
$\linebreak$
\vspace{-0.5in}
\hspace{0.30in} 
\end{figure}

Figure~\ref{fig:nn_data_bkg} shows  $D_{NN}$ for the 3,043 event data sample 
and  for background
normalized to the same number of events. Also shown for comparison is  the
expected
$t\overline{t}$~ Monte Carlo signal for $m_t =$ 180~GeV/$c^2$ multiplied 
by a factor of ten.
 The final event sample used in
the fit to the top mass was defined by a cutoff in $D_{NN}$, which was
chosen to minimize the uncertainty on the extracted top mass.
For a given cutoff on 
$D_{NN}$, and a given mass value $m_t$, 
a distribution was  
composed by adding the background 
mass distribution to the signal mass
distribution, with the signal normalized to the theoretical 
cross section~\cite{NNLOtheory}.
An ensemble of $\sim 100$ fake mass distributions was 
created by sampling from the combined 
distribution. The fitting procedure was applied to
each fake mass distribution to yield a fitted mass $m_{fit}$.
 We thereby
obtained a distribution of fitted masses, characterized by a mean and
an RMS, for the given $D_{NN}$ cutoff and the 
given value of $m_t$. The
procedure was repeated for different $D_{NN}$ cutoffs and 
for top quark mass values of 155, 165, 175, 185 and 195 GeV/$c^2$. 
We found that the cutoff $D_{NN} > 0.97$ 
minimizes the RMS in the fitted mass distributions for  
the five top quark masses considered.

\begin{figure}[hbt]
\vspace*{1.0cm}
\centerline{\epsfxsize=8.0cm\epsfysize=9.0cm\epsfbox{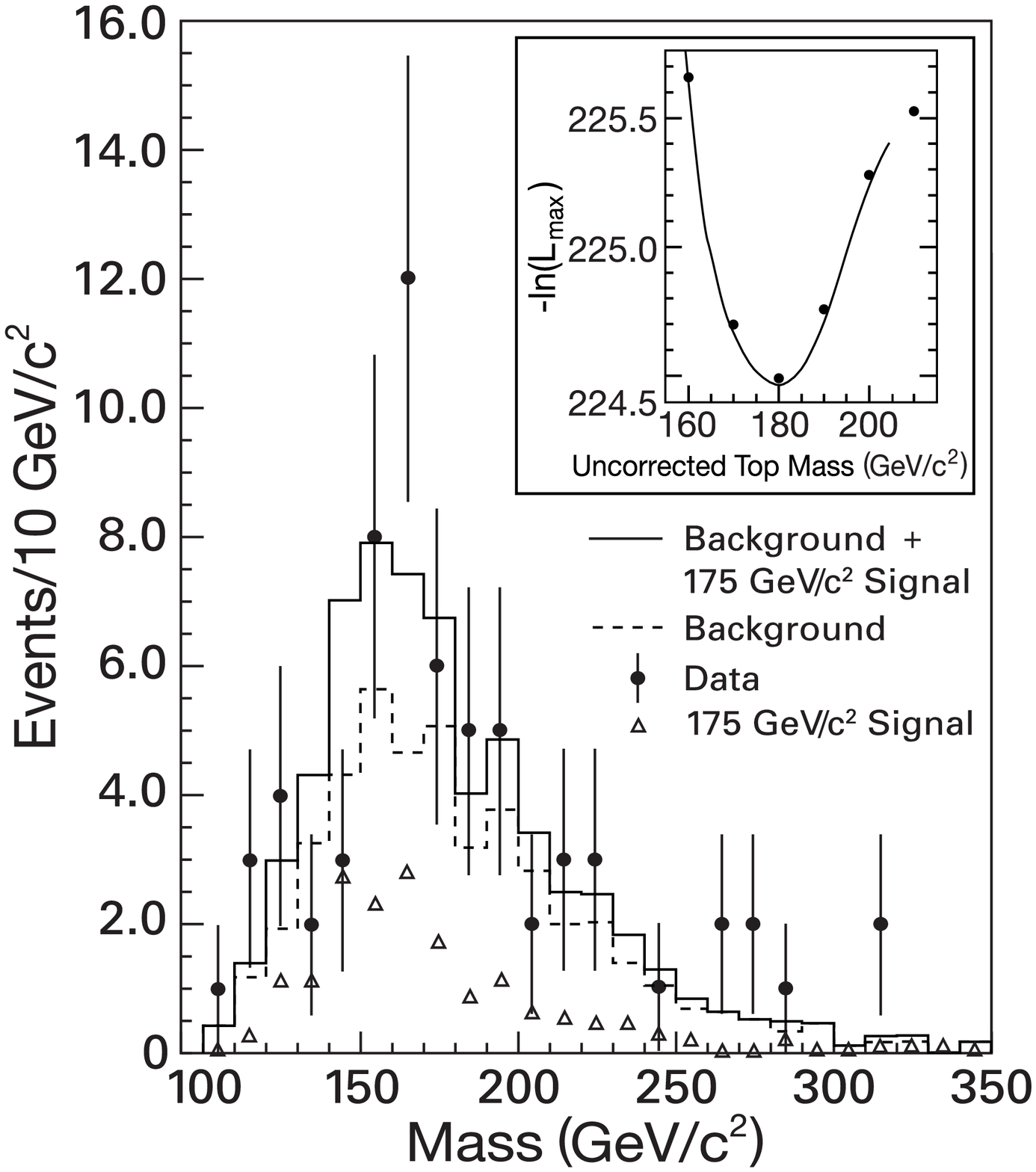}}
\caption{Data and  the sum of background and Monte Carlo signal plotted as a function of the  
mean mass, $M$. Insert is $-\ln{L_{max}}$ 
as a function of the top quark mass.}
\label{fig:data_bkg_nn_cut_v2}
\end{figure}

  When applied to the 3,043 events, the requirement of $D_{NN} > 0.97$ reduced
the dataset to a final sample of 65 events.  
Figure \ref{fig:data_bkg_nn_cut_v2} shows a comparison between
the observed mass distribution in the data
 and the sum of background and 175 GeV/$c^2$ top quark signal scaled to the
observed number of top events (see below). The fitting procedure, with
21 mass values, was
applied to the observed mass distribution to yield a mass estimate,
which was corrected for a  bias~\cite{thesis} of 2.6 GeV/$c^2$  using the relationship
\begin{equation}
\label{equation:calibration}
m_{fit} = 0.712 \,  m_t + 53.477 \, \, \, \mbox{GeV/$c^2$} \, ,
\end{equation}  
determined from the Monte Carlo studies.
The  uncertainties (also bias-corrected) 
were defined to be the 68\% interval about the minimum in the 
log-likelihood curve. 
A  systematic uncertainty of $5\%$  arises  from the discrepancy 
between the jet energy scale
in Monte Carlo simulations and  that in 
data~\cite{dzerocal}. 
The fitting of the tag-rate function introduces a normalization uncertainty of 
14.9 $\%$. The systematic uncertainty also receives a contribution from  the bin-by-bin uncertainty 
due to the limited number of untagged events used to model the background.
 The effect of these systematic 
uncertainties on the measured top quark mass
was obtained by repeating the fits  
varying the nominal values by their systematic uncertainties.
The effect of a small signal contribution  
to the background sample
was checked and found not to affect the determination of the top quark mass.

The insert in Figure \ref{fig:data_bkg_nn_cut_v2} shows the 
negative log-likelihood
as a function of the top quark mass for six points near the minimum. 
After bias correction, we find a top quark mass of
178.5 $\pm$ 13.7 ~(stat)~$\pm$ 7.7~(syst)~GeV/$c^2$.
As a consistency check of our measurement, given the measured top quark
mass, we can estimate the $t\overline{t}$~production cross section at
that mass.  The estimated signal in the 65-event sample is $16.6 \pm 7$ events.  
This corresponds to a total cross section of
$11 \pm 5$ pb, which is consistent with the measured D\O~  
$t\overline{t}$~production cross section of 
5.6~$\pm$ 1.4~(stat) $\pm$ 1.2~(syst) 
pb~ for a top quark mass of 
172.1  GeV/$c^2$~\cite{meenajim}. 

In summary, we have measured the mass of the top quark, 
using the purely hadronic decay modes
in  $t\overline{t}$~ events, to be
178.5 $\pm$ 13.7 ~(stat)~$\pm$ 7.7~(syst)~GeV/$c^2$.
This is in good agreement with top quark
mass measurements in other decay channels.

\section*{Acknowledgements}
%
We thank the staffs at Fermilab and collaborating institutions, 
and acknowledge support from the 
Department of Energy and National Science Foundation (USA),  
Commissariat  \` a l'Energie Atomique and 
CNRS/Institut National de Physique Nucl\'eaire et 
de Physique des Particules (France), 
Ministry of Education and Science, Agency for Atomic 
   Energy and RF President Grants Program (Russia),
CAPES, CNPq, FAPERJ, FAPESP and FUNDUNESP (Brazil),
Departments of Atomic Energy and Science and Technology (India),
Colciencias (Colombia),
CONACyT (Mexico),
Ministry of Education and KOSEF (Korea),
CONICET and UBACyT (Argentina),
The Foundation for Fundamental Research on Matter (The Netherlands),
PPARC (United Kingdom),
Ministry of Education (Czech Republic),
A.P.~Sloan Foundation,
and the Research Corporation.
%

\begin{thebibliography}{99}



\vskip 0.25cm



\bibitem{quigg}
C.~Quigg, Physics Today {50}  (1997) 20; P.C.~Bhat, H.B.~Prosper and S. 
~Snyder, Int. J. Mod. Phys.\ { A13} (1998)  5113; D.~Chakraborty, J.~Konigsberg, and D.
~Rainwater, Ann. Rev. Nucl. and Part. Sci.\ { 53}
(2003) 301.

\bibitem{ewstudies}
CERN LEP Electroweak Working Group, http://lepewwg.web.cern.ch/LEPEWWG/.

\bibitem{cdfalljets}
 F.~Abe {\sl et al.} (CDF Collaboration), Phys. Rev. Lett. {79} (1997) 1992.



\bibitem{cdfmass}
 F.~Abe {\sl et al.} (CDF Collaboration), Phys. Rev. D {63} (2001)  03200



\bibitem{nature}
 V.M.~Abazov {\sl et al.} (D\O\ Collaboration), 
 Nature 429 (2004) 638.
 


\bibitem{PDG}
Review of  Particle Physics, Phys. Rev. D {66} (2002) 010001.
 

\bibitem{dzeroalljets}
 S.~Abachi {\sl et al.} (D\O\ Collaboration), Phys. Rev. Lett.
83  (1999) 1908; S.~Abachi {\sl et al.}, Phys. Rev. D
 60  (1999) 012001.


\bibitem{dzerodet}
 S.~Abachi {\sl et al.} (D\O\ Collaboration),  Nucl. Instrum. Methods 
 in Phys. Res. A {338}
 (1994)  185.


\bibitem{dzero95}
 S.~Abachi {\sl et al.} (D\O\ Collaboration), Phys. Rev. D {52} (1995) 4877.


\bibitem{dzerojets}
 B.~Abbott {\sl et al.} (D\O\ Collaboration), Phys. Rev. D
{64} (2001) 032003.
 
\bibitem{dzerocal}R.~Kehoe, {\sl Proceedings of the $6^{th}$ International Conference
on Calorimetery in High Energy Physics}, Frascati, Italy, edited by A.
~Antonelli, S.~Bianco, A.~Calcaterrra and F.L.~Fabbri  
(World Scientific, River Edge, NJ, 1996) 349, FERMILAB-Conf-96/284-E.

\bibitem{dzeromass}
 B. ~Abbott {\sl et al.} (D\O\ Collaboration), Phys. Rev. D
{58}  (1998) 052001-1.
 

\bibitem{HERWIG}
G.~Marchesini {\sl et al.},  Com. Phys. Comm. {67}  (1992) 465. 


\bibitem{GEANT}
R.~Brun and C.~Carminati, 
``GEANT Detector Description and Simulation Tool,'' 
CERN Program Library Writeup W5013 (1993) (unpublished).


\bibitem{thesis}
B. Connolly, {\sl  ``Measurement of the Top Mass in the All-Jets Channel with the D0 Detector at
the Fermilab Tevatron Collider"}, Ph.D. thesis, Florida State University (2002), 
FERMILAB-THESIS-2002-22 (unpublished).


\bibitem{BPS}
P.C.~Bhat, H.B.~Prosper, and S.~Snyder, Phys. Lett. B {407} (1997) 73.


\bibitem{collider_physics}
J.D.~Bjorken and S.J.~Brodsky, Phys. Rev. D
{1} (1970) 1416;  V. D.~Barger and R.J. N.~Phillips, {\sl Collider Physics}
 (Addison-Wesley, Reading, MA, 1997) 280.

\bibitem{Tkachov}
This variable was inspired by discussions between
members of the D\O\ Collaboration and
Fyador Tkachov; F.~Tkachov, Int. J. Mod. Phys. A {12}  (1997) 5411.

\bibitem{NN}
V.~Rao and H.~Rao, {\sl C++ Neural Networks and Fuzzy Logic, 2nd Edition}, (MIS:Press,
New York, NY, 1995); E.K.~Blum and L.K.~Li, Neural Networks {4} (1991) 511;
D.W.~Ruck  {\sl et al.}, IEEE Trans. Neural Networks {1} (1990) 296.

\bibitem{NNLOtheory}
N. Kidonakis, Phys. Rev. D
{64} (2001) 014009.

\bibitem{meenajim}
V.M.~Abazov {\sl et al.} (D\O\ Collaboration), Phys. Rev. D
{67} (2003) 012004.

\end{thebibliography}
\end{document}